%% file: Main.tex
\documentclass[letterpaper,11pt]{article}

\input{Structure.tex}

\title{%
\vspace{-1em}
\centering\boldmath\Large\bfseries%
Dirac equation in curved spacetime: the role of local Fermi velocity%
\bigskip\bigskip}

\author[1]{B.\ Bagchi\footnote{\href{mailto:bbagchi123@gmail.com}{bbagchi123@gmail.com}}}

\author[2]{A.\ Gallerati\footnote{\href{mailto:antonio.gallerati@polito.it}{antonio.gallerati@polito.it}}}

\author[3]{R.\ Ghosh\footnote{\href{mailto:rg928@snu.edu.in}{rg928@snu.edu.in}}}

\affil[1]{%
Brainware University, Kolkata, Barasat, West Bengal, 700125, India%
\medskip
}

\affil[2]{% 
Politecnico di Torino, Dipartimento 
DISAT, corso Duca degli Abruzzi 24, 10129 Torino, Italy%
\medskip
}

\affil[ ]{%
\makebox[0em][c]%
{\centering \textsuperscript{3}Shiv Nadar Institute of Eminence, Physics Dept., Gautam Buddha Nagar, Uttar Pradesh 203207, India}%
\medskip
}

\date{}
\begin{document}
\maketitle

\vspace{-1.5em}
\begin{abstract}
We study the motion of charge carriers in curved Dirac materials, in the presence of a local Fermi velocity. An explicit parameterization of the latter emerging quantity for a nanoscroll cylindrical geometry is also provided, together with a discussion of related physical effects and observable properties.
\end{abstract}

\bigskip\medskip

%\vspace{-0.5em}
\noindent
\begin{tabularx}{\textwidth}{@{}r @{}X}
\textbf{Keywords: } & Dirac equation, graphene, local Fermi velocity, nanoscrolls.
\end{tabularx}
%\vspace{1em}

%\tableofcontents
%\pagebreak
%\newpage

\bigskip

\section{Introduction}
Dirac equation is one of the most relevant contributions in the history of quantum mechanics. Over the decades, its study has been conducted from different points of view \cite{thaller1992dirac,bjorken1964relativistic,Peskin:1995ev}, with countless applications in many areas of physics. The reformulation of the Dirac formalism in curved backgrounds is an appealing field of research due to its remarkable applications in high-energy physics, quantum field theory, analogue gravity scenarios and condensed matter.\par
A real, solid-state system where to observe the properties of Dirac spinorial quantum fields in a curved space is provided by graphene and other two-dimensional materials. The latter have attracted great interest because of their electronic, mechanical and optical characteristics \cite{novoselov2004electric,Novoselov:2005kj,Gusynin:2006ym,katsnelson2007graphene}.
The analysis of the electronic structure of these special substrates has received further contributions from theoretical studies on the dynamics of Dirac particles in curved spacetime. In particular, an effective low-energy model, formally equivalent to a Dirac equation, emerges in graphene-like materials; the resulting formulation then provides a powerful tool to describe the motion of the long-wavelengths charge carriers \cite{CastroNeto:2007fxn,geim2007rise,Cortijo:2006xs,geim2009graphene,Vozmediano:2010zz,Amorim:2015bga,Downing:2016pdl,Gallerati:2018dgm,fillion2021numerical,Gallerati:2021rtp,Hamil:2022yuc,deoliveira2022connecting}. Furthermore, from a practical perspective, stability, flexibility and near-ballistic transport at room temperature, make Dirac materials \cite{Vafek:2013mpa} an ideal framework for many nanoscale applications \cite{li2008chemically,stankovich2006graphene,bonaccorso2010graphene,sun2010graphene,lui2010ultrafast,Andrei:2012my}.
%\par

\paragraph{Curvature effects.}
In the large wavelength regime \cite{Peres:2010mx,Cortijo:2006mh,Cortijo:2006xs}, where quasiparticles wavelengths are large if compared with the honeycomb dimensions, curvature effects in Dirac materials have been extensively studied \cite{Kleinert1989gauge,CastroNeto:2007fxn,deJuan:2012hxm}. Intrinsic curvature deformations involve inelastic effects (usually coming from the formation of disclination-type defects), that turn out to work as potentials for pseudomagnetic fields modulated by curvature. Extrinsic curvature is an elastic effect coming from graphene sheet warping, determining a variation of the bond angles between orbitals and introducing peculiar effects in the electronic properties of the material \cite{kim2011multiply,fogler2010effect,Gallerati:2022egf}.\par
Effects from intrinsic curvature have been subject of much more attention within the graphene community, the inelastic deformation giving rise to the emergence of pseudogauge fields directly affecting the carrier dynamics and the electronic properties of the sample \cite{Vozmediano:2010zz}. In the following, instead, we focus on an extrinsic curvature framework, where the contributions of the pseudogauge fields are neglected and the physical properties are determined by the specific folding of the sample, expressed in terms of the related parametrization.

\paragraph{Local Fermi velocity.}
It is well recognized that a gap formation in graphene-like systems \cite{de2007charge,gui2008band} suggests the inclusion of a spatially-varying Fermi velocity in the sample \cite{de2013gauge,oliva2015generalizing,lima2015indirect,downing2017localization}. Confirmations in this direction also come from spectroscopy experiments \cite{deJuan:2012hxm,yan2013superlattice,hwang2012fermi,jang2014observation}. In addition, the need for a local Fermi velocity (LFV) emerges from the electronic transport in two-dimensional strained Dirac materials \cite{Phan:2021wry}; the latter feature has triggered a wide range of researches \cite{oliva2018fingerprints,oliva2020effective,Ishkhanyan:2018nlm,mustafa2013dirac,Ghosh:2021yly,Bagchi:2022sya,valencia2022position}.\par
From a theoretical point of view, the study of an effective Dirac behavior from lattice symmetry or, in a more general context, from low-energy expansion for distorted lattices in continuum approximation \cite{manes2007symmetry,park2009making} has allowed a better understanding of the correspondence between the lattice formulations and the quantum field covariant approach in curved space, justifying an emergent spatial dependence for the Fermi velocity \cite{deJuan:2012hxm}. On the other hand, the studies on heterostructures in graphene-like materials suggest that the dynamics of the charge carriers in the substrate can be described in terms of an effective Dirac model with a non-constant Fermi velocity \cite{peres2009scattering}. The latter features a position dependence that can be induced, e.g., by the presence of lattice strains, by specific curved geometries producing interlayer interactions, by deposition of graphene layers on suitable substrates giving rise to local on-site potentials spoiling the original symmetry \cite{gui2008band,oliva2015generalizing,yan2013superlattice,hwang2012fermi,giovannetti2007substrate,zhou2007substrate}.\par
A related, critical aspect in graphene-like materials is the generation of position-dependent mass (PDM) effects \cite{bagchi2004general,bagchi2005deformed,peres2006dirac,grassi2009tight,yannouleas2014beyond,reijnders2018electronic,Oliveira:2018xqt,david2012solutions,downing2019trapping,serafim2019position,schulze2020arbitrary,Schulze-Halberg:2020mil,schulze2022darboux,yesiltas2008new}. This has led to several theoretical studies, including those on wave packet dynamics and propagation in topological materials \cite{raghu2008analogs,bernevig2013topological,xie2019wave,hu2020linear}.
In the context of PDM, we are dealing with a generalized form of the Schr\"{o}dinger equation, involving a set of ambiguity parameters. In this regard, von Roos \cite{vonRoos:1983zz} showed that it is possible to construct a Hermitian form for the governing Hamiltonian, whose expression depends on the above-mentioned parameters.\par\medskip
In this work we consider some features of electron dynamics in Dirac materials, in the long-wavelength approximation and in the presence of a non-trivial local Fermi velocity. We focus, in particular, on  nanoscroll geometry. Graphene nanoscrolls are physically relevant due to their remarkable electronic, mechanical and optical properties, originating from their peculiar curved geometry \cite{zhang2011bending,xu2010geometry}. In this regard, we want to discuss physical observables  %predictable experimental properties,
exploiting a low-energy model with an effective Dirac equation for the charge carriers. 

As we will discuss, 
%in Sect.\ \ref{sec:model}  
the nanoscroll geometry is intrinsically flat: since the emerging local pseudomagnetic fields are proportional to the local Gaussian curvature \cite{Castro-Villarreal:2016wul,Morales:2022zws}, we do not consider them in our cylindrical geometry, the associated Gaussian curvature being zero everywhere.\par
The paper is organized as follows. In Sect.\ \ref{sec:curved}, we discuss the procedure we use to parameterize the underlying curved spacetime, restricting on a set of suitable cylindrical coordinates for the related Dirac equation. In Sect.\ \ref{sec:LFV}, we analyze the influence of the LFV on the Dirac equation, reformulating the problem in terms of modified $\gamma$-matrices. In Sect.\ \ref{sec:model}, we take into account the impact of PDM in our formulation and discuss the explicit case of a graphene nanoscroll, taking into account the density of states as a simple observable and commenting on the measurement issues. Finally, in Sect.\ \ref{sec:summary}, we briefly summarize the obtained results.

\section{Parameterizing the curved spacetime}
\label{sec:curved}
We start recovering the coordinates of a conventional three-dimensional spacetime in cylindrical coordinates as specified in \cite{Gallerati:2018dgm,Gallerati:2022egf}:
\begin{equation}
    x^\textsc{a} = (t, x, y, z) \quad\Rightarrow\quad 
    x^\mu = (t, \phi, z)\:,
\end{equation}
having expressed the former coordinates as
\begin{equation}
    x = r(\phi) \cos (\phi)\,, \qquad y = r (\phi) \sin (\phi)\,, \qquad z = z\,,
\end{equation}
with an explicit $\phi$--dependence for the radius.
Making use of the Jacobian ${J^\textsc{a}_\mu = \frac{\partial x^\textsc{a}}{\partial x^\mu}}$, we find the explicit form of the metric $g_{\mu\nu}$:
\begin{equation} \label{gmunu}
g_{\mu\nu} 
\,=\,\left(J^\textsc{a}_\mu\right)^\textsc{\!t}\eta_{\textsc{ab}}\:J^\textsc{b}_\nu
\:=\:\operatorname{diag} \left(1,\,-f(\phi)^2,\,-1\right)\,,
\end{equation}
where the quantity $f(\phi)^2$ reads
\begin{equation}
f(\phi)^2 = r(\phi)^2 + r'(\phi)^2\:.
\label{eq:fphi}
\end{equation}
As a result, the line element for the curved space assumes the form
\begin{equation}
ds^2 = g_{\mu\nu}\,dx^\mu \,dx^\nu = dt^2 - f(\phi)^2 \,d\phi^2 -dz^2\:.
\end{equation}
The flat-space Dirac equation is defined in terms of flat $\gamma$-matrices that we can parameterize in terms of Pauli matrices as
\begin{equation}
\gamma_a=\{ \sigma_3,\,-i\,\sigma_1,\,i\,\sigma_2 \}\,,\qquad 
a=0, 1,2    
\end{equation}
and obey the Clifford algebra
\begin{align} \label{flatclifford}
\{\gamma^a ,\gamma^b\}=2 \,\eta^{ab} \,\mathds{1}\,, 
\qquad\quad
\eta^{ab}= \begin{pmatrix}  1 & 0\\ 0 & -\mathds{1}
\end{pmatrix}.
\end{align}
To derive the form of the Dirac equation in curved space, we consider a tetrad basis at each spacetime point. The $\gamma_\mu$ matrices for the curved background, that we label by Greek indices, are obtained by means of the vielbeins $e_\mu{}^{a}$:
\begin{align} \label{curvedgamma1}
\gamma_\mu = e_\mu{}^a\,\gamma_a\,,
\qquad\quad
e_\mu{}^a = \operatorname{diag} \left(1,\,f(\phi),\,1\right)\,,
\end{align}
where $\gamma^\mu=g^{\mu\nu}\,\gamma_\nu$ satisfy the curved space Clifford algebra
\begin{align} \label{curvedclifford}
\{\gamma^\mu ,\gamma^\nu\}=2\,g^{\mu \nu} \,\mathds{1}, \qquad\quad \mu = 0, 1, 2.
\end{align}
The flat Dirac equation in the presence of an electromagnetic field $A^a(x)$,
\begin{align} \label{flatDiracconstvf}
    \left[i\hbar\, \gamma^a\left(\partial_a + \frac{g}{i\hbar}\,A_a\right)-m \right ] \Psi=0\:,
\end{align}
acquires, in the curved background, the form:
\begin{align} \label{curvedDiracconstvf}
    \left[ i\hbar\, \gamma^\mu\left(\partial_\mu +\Omega_\mu + \frac{ g}{i\hbar} A_\mu\right)-m \right] \Psi=0\:,
\end{align}
where $\Omega_\mu=\frac{1}{4}\,\omega_\mu{}^{ab}\,M_{ab}$ is the spin affine connection, expressed in terms of the spin connection $\omega_\mu{}^{ab}$ and Lorentz generators $M_{ab}=\frac{1}{2}[\gamma_a,\gamma_b]$ \cite{fock1929geometrization,parker2009quantum}.
According to the tetrad postulate, the gamma matrices of the curved space are covariantly constant. Furthermore, the existence of a confining potential is crucial in the thin-layer quantization approach %which makes the component arising due the presence of the Ricci scalar vanish
\cite{liang2018pseudo,wang2018geometric}.\footnote{%
This leads to two parts of the Dirac equation: one Schrödinger-like equation with an externally applied electromagnetic field, and another expression corresponding to the spin-orbit interaction term on the curved surface. We are interested in the former contribution only.
}
\par
Imposing the cylindrical symmetry of space, the spin affine connection $\Omega_\mu$ vanishes and \eqref{curvedDiracconstvf} simplifies to
\begin{align}
    \left[ i\hbar\, \gamma^\mu\left(\partial_\mu + \frac{ g}{i\hbar} A_\mu\right)-m \right] \Psi=0\:,
\end{align}
having the same form of \eqref{flatDiracconstvf}, but expressed in terms of the curved $\gamma^\mu$ matrices.

\section{Impact of a local Fermi velocity}
\label{sec:LFV}
For heterostructures, the Fermi velocity becomes position-dependent \cite{peres2009scattering,concha2010effect,panella2012bound}: 
\begin{equation}
\vF\;\rightarrow\; \vF(\phi,z)\:,
\end{equation}
and this also affects Dirac equation \eqref{flatDiracconstvf}, that is modified as \cite{peres2009scattering,raoux2010velocity,krstajic2011ballistic}
\begin{align} \label{curvedDiracLFV1}
    \left[ i\hbar\, \Gamma^a  \sqrt{\vF} \left( \partial_a \sqrt{\vF} +  \frac{g}{i\hbar}\, \sqrt{\vF}\,A_a \right)-m \right] \Psi=0\:,
\end{align}
where 
\begin{equation}
\Gamma^a = \left\{\frac{\gamma^0}{\vF(\phi,z)},\,\gamma^1,\,\gamma^2\right\}\,
\label{eq:tilgamflat}.    
\end{equation}
For convenience, we also introduce the $\Gamma^\mu$ matrices 
\begin{align}
\Gamma^\mu=\left\{\frac{1}{\vF(\phi,z)}\,e^{0a}\,\gamma_a,\:e^{1a}\,\gamma_a,\:e^{2a}\,\gamma_a\right\},\qquad\quad
e^{\mu a}=g^{\mu\nu}\,e_\nu{}^a\,,\quad
\label{eq:tilgamcurv}
\end{align}
so that a curved space Dirac equation in the presence of LFV takes the form
\begin{align}
    \left[ i\,\Gamma^\mu \sqrt{\vF} \left( \partial_\mu \sqrt{\vF} + \frac{g}{i}\, \sqrt{\vF}\,A_\mu \right)-m \right] \Psi=0\:,
\end{align}
where, from now on, we assume $\hbar=1$.\par\smallskip
Let us now study more in detail eq.\ \eqref{curvedDiracLFV1}.
In order to simplify our analysis, we define the field $\chi(\phi,z)$:
\begin{align}
\sqrt{\vF(\phi,z)}\,\Psi(\phi,z) =\chi(\phi,z)\:,
\end{align}
so that the above equation can be rewritten as
\begin{align} \label{curvedDiracLFV}
   \left[ i\,\Gamma^\mu\,\vF \left(\partial_\mu  + \frac{ g}{i} \,A_\mu \right)-m \right] \chi=0\:.
\end{align}
%
% We can also represent the above equation by
% %
% \begin{align}
%    \left[ i \bar{\gamma}^\mu \left( \partial_\mu  + \frac{ g}{i}  A_\mu \right)-m \right] \chi=0
% \end{align}
% %
% where $\bar{\gamma}^\mu= \{ \Gamma^0,\, \vF(z)\,\Gamma^j \}$, obeying the Clifford algebra
% $\{\bar{\gamma}^\mu, \bar{\gamma}^\nu\}=2 \,\bar{g}^{\mu\nu}$.
% Explicitly, the components of $\gamma_\mu$ read
%
%\begin{align}
%     \gamma_a &=\{ \sigma_3, -i\sigma_1, i\sigma_2 \} = \left\{
%     \begin{pmatrix}    1 & 0\\   0 & -1  \end{pmatrix},
%     \begin{pmatrix}    0 & -i\\   -i & 0  \end{pmatrix},
%     \begin{pmatrix}    0 & 1\\   -1 & 0  \end{pmatrix} \right\}\,,
% \\[2.5ex]
% \gamma_\mu &= e_\mu {}^a\,\gamma_a = \left\{
%     \begin{pmatrix}    1 & 0\\   0 & -1\end{pmatrix},
%     \begin{pmatrix}    0 & i f(\phi)\\   i f(\phi)& 0  \end{pmatrix},
%     \begin{pmatrix}    0 & -1 \\   1 & 0  \end{pmatrix} \right\}\,,
% \end{align}
%
Explicitly, the components of $\Gamma^\mu$ read
\begin{align}
    \Gamma^\mu= g^{\mu\nu}\,\Gamma_\nu  =\left\{
    \frac{1}{\vF} \begin{pmatrix}    1 & 0\\   0 & -1\end{pmatrix},
    \begin{pmatrix}    0 & -\frac{i}{f(\phi)}\\   -\frac{i}{f(\phi)}& 0  \end{pmatrix},
    \begin{pmatrix}    0 & 1 \\   -1 & 0  \end{pmatrix} \right\}\,,
\end{align}
where $f(\phi)$ is given by \eqref{eq:fphi}, while we have considered the inverse of \eqref{gmunu} for $g^{\mu\nu}$ and employed \eqref{eq:tilgamflat} and \eqref{eq:tilgamcurv}.

\subsection{Dirac equation}
Using the explicit form of the $\Gamma^\mu$ matrices, we can write equation  \eqref{curvedDiracLFV} in the form
\begin{align}
    \begin{pmatrix}
  E-m & \frac{\vF}{f} \left(\partial_\phi+\frac{g}{i}A_\phi \right) + i \vF\left(\partial_z+\frac{g}{i}A_z\right) 
  \\[1ex]
  \frac{\vF}{f} \left(\partial_\phi+\frac{g}{i}A_\phi \right) - i \vF\left(\partial_z+\frac{g}{i}A_z\right) & -(E+m)
\end{pmatrix}  \begin{pmatrix}
\chi_1 \\ \chi_2
\end{pmatrix} =0\:,
\label{eq:DiracLFV}
\end{align}
where we adopted the column representation for $\chi=\left(\chi_1,\, \chi_2 \right)$ and restricted the local Fermi velocity to be a function of $z$ only, $\vF=\vF(z)$.\par\smallskip
We now make the following ansatz for $\chi$:
\begin{align}
\chi=e^{-i\,\lambda\,E\,t}\begin{pmatrix} 
\chi_1(\phi,z) \\[\jot] \chi_2(\phi,z) \end{pmatrix}=
e^{-i\,\lambda\,E\,t} \;e^{i\,k_\phi \mathlarger{\int}^{^\phi}\!\!\!f(\phi')\,d\phi'}
\begin{pmatrix}
\chi_{+} (z) \\ \chi_{-} (z)
\end{pmatrix},
\end{align}
were both $\chi_{+}$ and $\chi_{-}$ are functions of $z$ only. We have also assumed the vector $A_\mu$ to be expressed as
\begin{equation}
A_\mu=\big(0,\,A_\phi(\phi,z),\,A_z(\phi,z)\big)\,.
\end{equation}
%
%where $A_\phi(z)$ and $A_z(z)$ are function of $z$ only. 
Expanding then \eqref{eq:DiracLFV}, we obtain a pair of coupled equations of the form
\begin{align} \label{uncoupledCurvedDirac}
 \left[ \frac{\vF}{f} \left(\partial_\phi+\frac{g}{i}\,A_\phi \right) + i\, \vF\left(\partial_z+\frac{g}{i}\,A_z\right)\right] \chi_{2}+ (E-m)\, \chi_{1} =0\:,
 \\[1.75ex]
 \left[\frac{\vF}{f} \left(\partial_\phi+\frac{g}{i}\,A_\phi \right) - i \,\vF\left(\partial_z+\frac{g}{i}\,A_z\right)\right] \chi_{1} -(E+m) \,\chi_{2} =0\:.
 \end{align}
Uncoupling the above system, we get two separate second-order differential equations, corresponding to separate conditions for the upper component $\chi_+$ and the lower component $\chi_-$\,. In particular, we find for $\chi_+$
\begin{equation}
\begin{aligned}[b]
\Big[&\vF^2\,\partial^2_z+ \left( \vF\, \vF'+2\,\vF^2\, \frac{g}{i}\,A_z \right) \partial_z +\left(g\,\vF\, (\vF\,W_\phi)'- k_\phi\,\vF\,\vF' + 2\,g\,k_\phi\,\vF^2\,W_\phi - g^2 \,\vF^2\,W_\phi^2 - k_\phi^2\, \vF^2 \right)+  
\\[\jot]
&+\left(\frac{g}{i}\,\vF\,(\vF\,A_z)'- g^2\,\vF^2\, A_z^2 \right) + \left(E^2-m^2\right) \Big]\;\chi_+ =0   \:,
\end{aligned}   
\label{eq:chiplus0}
\end{equation}
where we have also used the separation $A_\phi=f(\phi)\,W_\phi(z)$.\par
In order to get an exact solution we set $A_z=0$ which modifies \eqref{eq:chiplus0} to the form
\begin{equation}
\begin{aligned}[b]
\Big[&\vF^2\,\partial^2_z+ \vF\,\vF'\,\partial_z +\Big(g\,\vF\, (\vF\,W_\phi)'- k_\phi\,\vF\,\vF' + 2\,g\,k_\phi\,\vF^2\,W_\phi -g^2\,\vF^2\,W_\phi^2 - k_\phi^2\,\vF^2 \Big)+
\\[\jot]
&+\left(E^2-m^2\right) \Big]\;\chi_+ =0\:.
%\\ 
% \Big[ \vF^2\,\partial^2_z+ \vF\,\vF'\,\partial_z +\Big( k_\phi\,\vF\,\vF'-g\,\vF\, (\vF\,W_\phi)' + 2\,g\,k_\phi\,\vF^2\,W_\phi -g^2\,\vF^2\,W_\phi^2 - k_\phi^2\,\vF^2 \Big) \\
% +\left(E^2-m^2\right) \Big]\;\chi_- =0\:.
\end{aligned}
\label{eq:chiplus}
\end{equation}
Note that this ansatz removes the imaginary term in \eqref{eq:chiplus0}. The correspondent equation for $\chi_{-}$ can be found by substituting \,$k_\phi\to-k_\phi$\, and \,$g\to-g$\, in the above expression.\par
Since we want explicit analytic solutions, we need to define the explicit form of the LFV and the vector component of $A_\phi$\,. In the following section, we exploit a typical choice. We also briefly comment PDM schemes characterized by an extended Schr\"{o}dinger equation, the latter in turn dependent on a generalized class of potentials containing a set of ambiguity parameters.

\section{An explicit model}
\label{sec:model}
The study of position-dependent mass problems \cite{vonRoos:1983zz} has found a huge amount of applications in the last two decades. These include the study of the electronic properties of condensed-matter systems, such as compositionally-graded crystals, quantum dots and liquid crystals \cite{bastard1990wave,de2022gaussian}, quantum many-body systems focusing on the nuclei, quantum liquids, helium and metal clusters \cite{ring2004nuclear}. 
Furthermore, PDM has found relevance in different theoretical contexts that include supersymmetric quantum mechanics \cite{Bagchi2008generalized,bagchi2005deformed,Quesne:2018ppp}, coherent states \cite{Cruz2009position} and use of indefinite effective mass \cite{znojil2012schrodinger}. Several techniques have been developed to tackle PDM problems like the so-called point canonical transformation \cite{quesne2009point} and potential algebra formalism (see, for instance, \cite{Bagchi:2022sya} and references therein). \par\smallskip
The feature of spatial modulations of Fermi velocity leading to velocity barriers is an interesting field of research (see for example \cite{deJuan2007charge}). Several tractable forms of LFV, facilitating closed form solutions of the governing system, have already been envisaged \cite{downing2017localization,mustafa2013dirac}. Of particular interest is the study of a certain range of models concerning the shape of the velocity profile, as well as the way in which the latter influences two-dimensional Dirac materials, in order to study localization effects and induced bound states. One such form involved the exponential velocity profile \cite{downing2017localization}, the role of which was explored on the problem of bound modes propagation along a waveguide, a situation of great interest where to realize electron optics
based on ballistic guiding in Dirac materials. Motivated by the latter \cite{downing2017localization} and by earlier studies on the PDM issue \cite{mustafa2013dirac,Ghosh:2021yly,Bagchi:2022sya}, we consider the following forms for the LFV and $W_\phi$ function: 
\begin{equation}
\vF(z)=v_0\,e^{\alpha z}\,,\qquad
W_\phi(z)=w_0\,e^{-\alpha z}\,,
\label{eq:vFz_Wz}
\end{equation}
where $v_0$ is the minimal Fermi velocity, found at the center of the barrier, $\alpha>0$ is related to the length scale of the problem and $w_0$ is a real constant.\par
\sloppy
%A few remarks about the above parametrization are in order. 
The search for exact solutions of Schr\"{o}dinger-like equations, in an extended framework with a PDM, has received several contributions where an exponential form has been proposed for the PDM itself \cite{dos2021probability,dong2022exact,gonul2002exact}. Going deeper into the role of the LFV, an interesting connection between the mass function and the form of the local velocity was also noted 
%${m(x)\leftrightarrow\vF(x)^{-2}}$ 
\cite{Ghosh:2021yly,mustafa2013dirac,Valencia-Torres:2022fiv}. 
%This immediately leads to the realization of the exponential character of the latter. 
The nature of the LFV, in the special framework of a $\mathrm{so}(2, 1)$ algebra, has recently been discussed in connection with a PDM-dependent Dirac equation in \cite{Bagchi:2022sya}.\par\smallskip
Using above ansatz \eqref{eq:vFz_Wz}, we see from \eqref{eq:chiplus} that the upper component $\chi_+$ satisfies
\begin{equation}
\Big[v_0^2\,e^{2\alpha z}\,\partial^2_z + \alpha\,v_0^2 \,e^{2\alpha z}\,\partial_z - \left(\alpha\,k_\phi\,v_0^2+k^2_\phi\,v_0^2\right)\,e^{2\alpha z} + 2\,B\,k_\phi\,v_0\,e^{\alpha z} - \kappa^2 \Big]\, \chi_+(z) =0\,, 
%\\
% \Big[v_0^2\,e^{2\alpha z}\,\partial^2_z + \alpha\,v_0^2 \,e^{2\alpha z}\,\partial_z - \left(-\alpha\,k_\phi\,v_0^2+k^2_\phi\,v_0^2\right)\,e^{2\alpha z} + 2\,B\,k_\phi\,v_0\,e^{\alpha z} - \kappa^2 \Big]\, \chi_-(z) =0\,,
\label{eq:chiplusLFV}
\end{equation}
where
\begin{equation}
\kappa^2= m^2 + B^2 -E^2\,,\qquad\quad B = g\,v_0\,w_0\,.\quad
\end{equation}
If we now make the change of variable
\begin{equation}
y=\int^z\!\frac{1}{\vF(q)}\,dq = -\frac{e^{-\alpha z}}{v_0 \,\alpha}\:,
\end{equation}
equation \eqref{eq:chiplusLFV} can be recast as
\begin{equation}
\left[\partial^2_y - \frac{k_\phi}{\alpha}\left(1+\frac{k_\phi}{\alpha}\right)   \frac{1}{y^2} - \frac{2\,B\,k_\phi }{\alpha}\:\frac{1}{y} - \kappa^2 \right] \chi_+(y) =0\,, \qquad -\infty <y< 0\:,\quad 
% \\
% \left[\partial^2_y - \frac{k_\phi}{\alpha}\left(-1+\frac{k_\phi}{\alpha}\right)   \frac{1}{y^2} - \frac{2\,B\,k_\phi }{\alpha}\:\frac{1}{y} - \kappa^2 \right] \chi_-(y) =0\,, \qquad -\infty <y< 0\:.\quad
\label{eq:chiplusLFVy}
\end{equation}
while the correspondent equation for $\chi_{-}$ is found through substitutions $k_\phi\to-k_\phi$ and $B\to-B$.\par
The solution of above \eqref{eq:chiplusLFVy} is given by
\begin{equation}
\chi_+(y) \;\propto\; \mathcal{M}\!\left(-\frac{B\,k_\phi}{ \alpha\,\kappa}\,,\, \frac{2\,k_\phi+\alpha}{2\,\alpha}\,;\:2\,y\,\kappa \right)\,,
\end{equation}
where $\mathcal{M}(a,b;y)$ is the well known Whittaker function.%
\footnote{%
The function $\mathcal{M}(a,b;z)$ is a solution of a Whittaker equation and is expressed in terms of the Kummer confluent hypergeometric function as $\mathcal{M}(a,b;z)=e^{-\frac{z}{2}}\,e^{b+\frac{1}{2}}\:{}_{1}F_{1}\!\left(b-a+\frac{1}{2},\,1+2\,b;\,z\right).$
}
In terms of the variable $z$, we then find
\begin{align}
\chi_+(z) =C_1\:\mathcal{M}\!\left(-\frac{B\,k_\phi}{ \alpha\,\kappa}\,,\,\frac{2\,k_\phi+\alpha}{2\,\alpha}\,;\:-\frac{2\,\kappa}{v_0\,\alpha}\, e^{-\alpha z} \right)\,.
\label{eq:solchiplus}
\end{align}
On the other hand, the lower component $\chi_-$ of the spinor wavefunction reads
\begin{align}
\chi_{-}(z) =C_1\:\mathcal{M}\!\left(-\frac{B\,k_\phi}{ \alpha\,\kappa}\,,\,\frac{-2\,k_\phi+\alpha}{2\,\alpha}\,;\:-\frac{2\,\kappa}{v_0\,\alpha}\, e^{-\alpha z} \right)\,.
\label{eq:solchimin}
\end{align}
For the energy levels, we exploit the relation
\begin{align}
\kappa_n = -\frac{B\, k_\phi}{k_\phi+\alpha + n\,\alpha}\:, \qquad n = 0,1,2,\dots
\end{align}
to determine
\begin{align}
E_n^2=m^2+B^2\left(1-\frac{k_\phi^2}{(k_\phi+\alpha +n\,\alpha)^2}\right)\,, \qquad n = 0,1,2,\dots
\end{align}
the above expression ensuring that we always get real energies.
%
%Observe that on increasing $n$ will decrease their values. %Furthermore, for the massless Dirac equation will correspond to simply $E = \pm gv_0w_0$.

\subsection{Application: density of states}
Once obtained an explicit solution $\chi$ to modified Dirac equation \eqref{eq:DiracLFV}, we can consider the probability density $\mathcal{P}=\sqrt{\det(g_{\mu\nu})}\:\chi^2$\,, in terms of which a normalization condition can be written as
\begin{equation}
\int d\Sigma\;\PP(\phi,z)
    =\int{\!dz\,d\phi\,\sqrt{r(\phi)^2+r'(\phi)^2}\,\abs{\chi(\phi,z)}^2}=1\;.
\end{equation}
%.
In order to write an expression for $\PP$, we need an explicit surface parameterization $r(\phi)$ for the sample surface, together with the previous solutions \cref{eq:solchiplus,eq:solchimin}.\par

\subsection{Nanoscrolls}
Graphene nanoscrolls \cite{xie2009controlled,braga2004structure,kim2011multiply} have received great attention due to their unusual properties and potential applications \cite{chen2007structural,mpourmpakis2007carbon,berman2015macroscale,
li2018superior,saini2021low}. They consist in carbon--based structures obtained by rolling a graphene layer into a cylindrical geometry \cite{xie2009controlled}, and exhibit some exciting 
%electronic and mechanical 
qualities due to their distinctively different conformation \cite{zhang2011bending,xu2010geometry}, making them conceptually interesting and experimentally relevant.\par
Unusual electronic and optical properties of carbon nanoscrolls are due to their unique topology and peculiar structure \cite{fogler2010effect,Gallerati:2018dgm,Gallerati:2022egf}. The
nanoscroll geometry is intrinsically flat, as shown by its vanishing Riemann components and Gaussian curvature. In this regard, emerging local pseudomagnetic fields $\mathcal{B}_\text{s}$ due to curvature effects are proportional to the local Gaussian curvature $\mathcal{R}$ \cite{Castro-Villarreal:2016wul,Morales:2022zws}, that in the cylindrical nanoscroll geometry vanishes.\par
On the other hand, nanoscrolls exhibit non-zero extrinsic curvature $\mathcal{K}_{ab}$, resulting from their non-trivial embedding in $\mathbb{R}^3$. The sample extrinsic curvature affects the experimental observables, giving rise to physical measurable effects \cite{CastroNeto:2007fxn,kim2011multiply,Gallerati:2018dgm,Gallerati:2022egf}. 
%, including an affection of the local number density. 
In this regard, the long-wavelength approximation also helps to narrow the field to charge carriers that are more likely to experience the extrinsic global curvature of the sample, justifying at the same time a continuum approximation for the substrate.\par\smallskip
The nanoscroll geometry can be explicitly parameterized in terms of Archimedean--type spirals in cylindrical coordinates \cite{chen2022formation,trushin2021stability,hassanzadazar2016electrical}. 
Let us then consider the following parameterization for the cylindrical geometry of the layer surface \cite{Gallerati:2018dgm,Gallerati:2022egf}:
\begin{equation}
r(\phi)=R\,{\left(1-\frac{1}{2\pi}\,\frac{\phi}{D N}\right)}^{\!2}\;,
\end{equation}
where the coefficient $D$ controls the distance between the layers and $R$ is a typical dimension for the radius of the cylinder. The integer number $N$ takes into account the windings of the wrapped graphene layer; from a physical point of view, it acts as a momentum cutoff, correctly restricting the analysis to the long-wavelength continuum approximation.%
\footnote{%
This requires that the number $N$ is small when compared with the ratio of the cylinder circumference to the graphene lattice dimension $a$\, $\left(N\ll \frac{2\pi R}{a}\right)$\:.
}
The boundary conditions in the cylindrical symmetry give rise to a quantization condition on the transverse component of the charge carriers momentum $k_\phi$ of the form \cite{Gallerati:2018dgm,Gallerati:2022egf}
\begin{equation}
k_\phi = \frac{2\pi}{\zetaN}\:,
\label{eq:kquantcond}
\end{equation}
having defined the geometrical parameter $\zetaN$
\begin{equation}
\zetaN~\equiv~
\int\limits^{2\pi N}_{0}{\!\!d\phi\:f(\phi)
%\sqrt{r(\phi)^2+r'(\phi)^2}
}\;,
\end{equation}
roughly expressing a measure of the cylinder spiral.\par
\begin{figure}[H]
\centering
\includegraphics[width=0.75\textwidth,keepaspectratio]{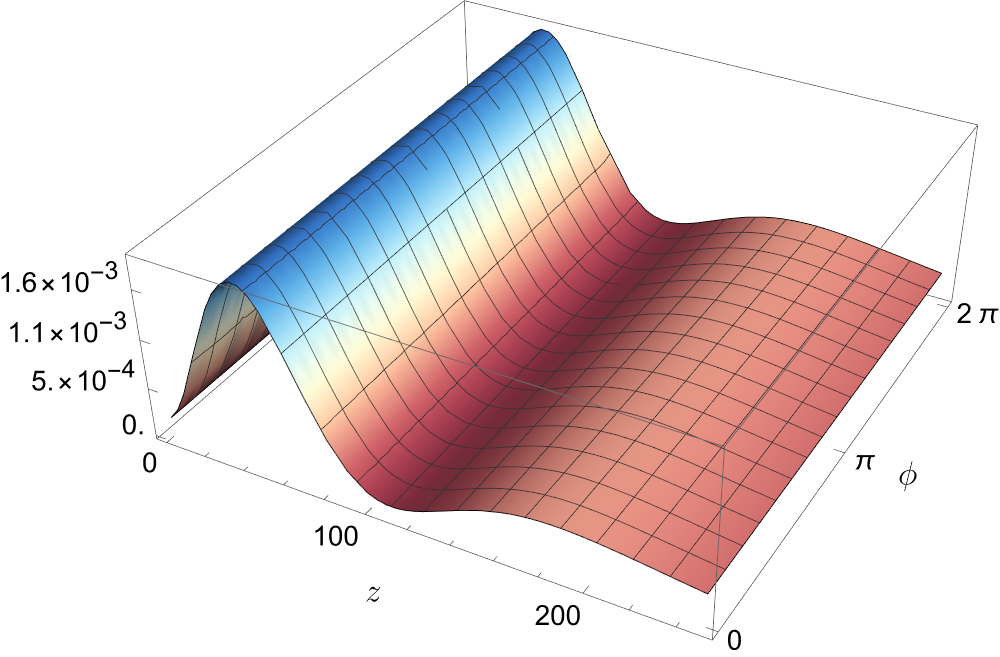}
\caption{Normalized probability density $\PP$ as a function of the coordinates $(\phi,\,z)$ for a nanoscroll geometry with parameters $N=4$, \,$R=15$, \,$D=100$, \,$g=10^{-3}$, \,$\alpha=1.5\times10^{-2}$, \,$v_0=50$, \,$w_0=100$\,.}%
\label{fig:prob_dens_N4}
\end{figure}

\begin{figure}[H]
\centering
\includegraphics[width=0.75\textwidth,keepaspectratio]{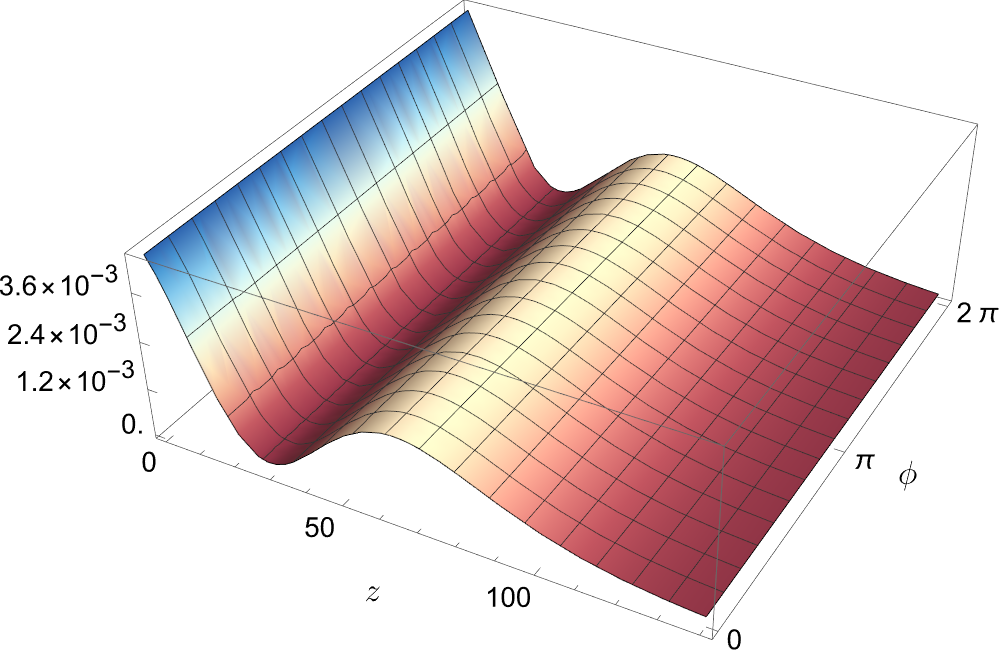}
\caption{Normalized probability density $\PP$ as a function of the coordinates $(\phi,\,z)$ for a nanoscroll geometry with parameters $N=5$, \,$R=5$, \,$D=100$, \,$g=10^{-3}$, \,$\alpha=2\times10^{-2}$, \,$v_0=200$, \,$w_0=100$\,.}%
\label{fig:prob_dens_N5}
\end{figure}

\begin{figure}[H]
\centering
\includegraphics[width=0.75\textwidth,keepaspectratio]{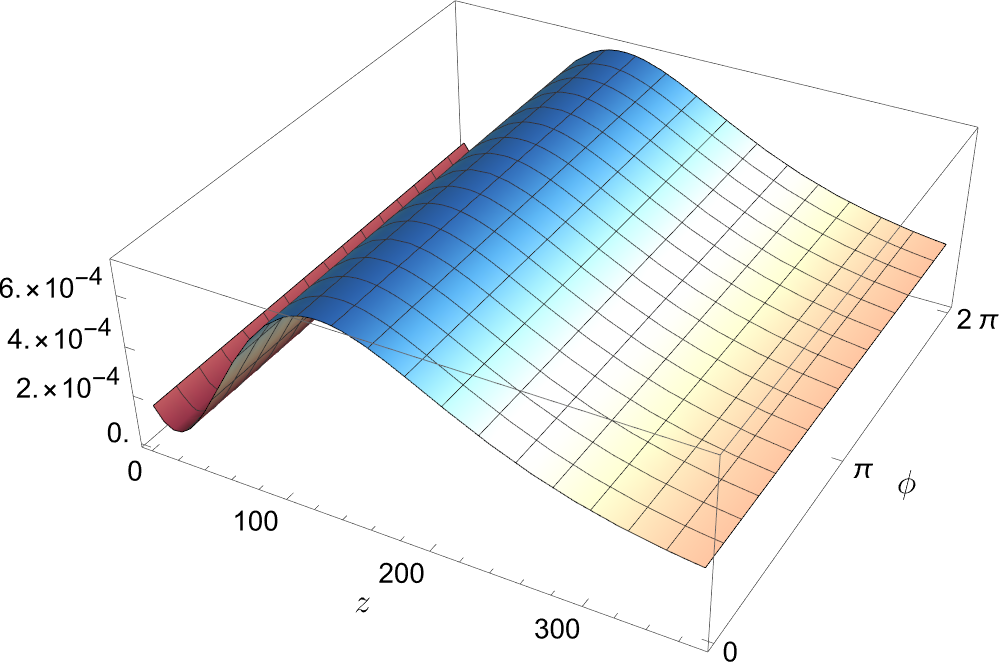}
\caption{Normalized probability density $\PP$ as a function of the coordinates $(\phi,\,z)$ for a nanoscroll geometry with parameters $N=3$, \,$R=30$, \,$D=100$, \,$g=10^{-3}$, \,$\alpha=10^{-2}$, \,$v_0=90$, \,$w_0=15$\,.}%
\label{fig:prob_dens_N3}
\end{figure}

\medskip
In \Cref{fig:prob_dens_N4,fig:prob_dens_N5,fig:prob_dens_N3} we plot the normalized probability density $\PP$ as a function of the coordinates $(\phi,\,z)$ for a nanoscroll geometry with different dimensions, number of windings $N$ and energy parameters.

\medskip
%\bigskip

\bigskip

\bigskip

We can see that the structure of the probability density function is weakly affected by displacements along the $\phi$ coordinate, while its variation is non-trivial along the $z$ direction, in agreement with our choice \eqref{eq:vFz_Wz}.

\subsection{Local density of states}
Given a location $X$ on the layer surface and an energy value $E$, the local density of states (LDOS) of the sample can be expressed in terms of the probability density as \cite{Chen1993introduction}
\begin{equation}
\LDOS(E,X,0)=\frac{1}{\varepsilon}\:\sum_{E-\varepsilon}^{E}\PP(X)\:,
%\qqquad\varepsilon\to0
\label{eq:LDOS}
\end{equation}
%
%in terms of the amplitude $\PP$.
for small values of $\varepsilon$, where the $0$--coordinate states that we are considering pseudoparticles on a two-dimensional space (zero-distance from the substrate surface). The LDOS then expresses the number of charge carriers per unit surface and unit energy range of size $\varepsilon$, at a given surface location $X$ and energy $E$. The sample LDOS is not only an interesting direct physical observable, but also a substrate feature of great importance for electronic applications, being the availability of empty valence and conduction states (states below and above the Fermi level) crucial for the transition rates.

\paragraph{Measurements.}
The sample LDOS can be mapped using a scanning tunneling microscope (STM). The latter is an experimental device based on quantum mechanical tunneling, where the wave-like properties of charge carries allow them to penetrate, through a potential barrier, into regions that are forbidden to them in the classical picture. STM spectroscopy provides insight into the surface electronic properties of the substrate, being the tunneling current strongly affected by the local density of states $\LDOS$. The latter is in turn related to the probability density $\PP$ through definition \eqref{eq:LDOS}.\par
A typical STM device consists of sharp conductive tip, brought within tunneling distance ($<\mathrm{nm}$) from a sample surface.
A small voltage bias $V$ is then applied between the probe tip and the substrate, causing charge carriers to tunnel across the gap, resulting in a tunneling current between the sample and the tip.
\footnote{%
To simplify our discussion, we are assuming that both materials have the same Fermi level.}\par
In the STM--map setup, the density of states at some fixed energy is mapped as a function of the position $(\phi,\,z)$ on the sample surface. If we assume $\varepsilon=e\,V$ to be very small with respect to the work function $\Phi_\textsc{w}$ (minimum energy required to extract an electron from the surface), the sample states with energy lying between $\EF-\varepsilon$ and $\EF$ are very close to the Fermi level and have non-zero probability of tunneling into the tip. The resulting tunneling current $\I$ is directly proportional to the number of states on the substrate within our energy range of width $\varepsilon$, this number depending on the local properties of the surface. Including all the sample states in the chosen energy range, the measured tunneling current can be modeled, in first approximation, as \cite{Chen1993introduction}
\begin{equation}
\I~\propto\,\sum_{\EF-e\,V}^{\EF}\PP\;e^{-2\,\Lambda\,d}\;,
\end{equation}
where $\Lambda$ is some decay constant in the barrier separation depending on $\Phi_\textsc{w}$\,. The exponential function gives the suppression for charge carriers tunneling in the classically forbidden region of width $d$ (sample-tip separation).
The tunneling current can be then measured, for constant separation $d$, at different $X$ positions and, for sufficiently small $V$, it can be conveniently expressed in terms of the LDOS of the sample as \cite{Chen1993introduction}:
\begin{equation}
\I(X)~\propto~\,\LDOS(\EF,X,0)\;e^{-2\,\Lambda\,d}\;e\,V\;.
\end{equation}
For finite bias voltage and different Fermi levels for the sample and tip, the tunneling current and its relation with the sample local density of states can be obtained from Bardeen time-dependent perturbation approach \cite{Chen1993introduction}.

\section{Summary}
\label{sec:summary}
The study of graphene-like systems is a captivating and multidisciplinary field of research, combining notion and techniques from quantum mechanics, general relativity and condensed matter physics. These special materials realize the physics of Dirac fermions in a real laboratory framework, the substrate acting as a lower-dimensional curved spacetime for the charge carriers. This provides a direct connection between condensed matter
and theoretical quantum models, with the possibility to explore the analogues of many high energy physics effects in a solid-state system \cite{Cvetic:2012vg,Stegmann:2015mjp,Alvarez:2011gd,Sepehri:2016nuv,Franz:2018cqi,Capozziello:2018mqy,Pedernales:2017eue,Kandemir:2019zyt,Andrianopoli:2019sip,Gallerati:2021htm,morresi2020exploring,Capozziello:2020ncr}.\par
In this paper we studied the governing equations of the charge carriers in Dirac materials, in the presence of a local Fermi velocity and non-trivial energy-mass parameters. The curvature of the sample mimics a curved spacetime background for the Dirac pseudoparticles, whose dynamics is then governed by a suitably modified Dirac equation. The procedure led us to an analytic expression for the wave functions of the quasiparticle modes, which was then applied to an explicitly example involving a nanoscroll geometry. We also discussed the impact on the sample density of states as a simple and straightforward physical observable.  

%\bigskip

\newpage

\paragraph{Acknowledgments.}
BB thanks Shiv Nadar Institute of Eminence for the initial support of this work. BB also thanks Mr.\ Phalguni Mookhopadhayay, Chancellor Brainware University, for constant encouragement. RG thanks Shiv Nadar University for the grant of senior research fellowship.
AG thanks prof.\ Francesco Laviano that
supported these studies with his funds.

\medskip

\paragraph{Data availability statement.}
All data supporting this study are included in the article.

\medskip

\paragraph{Conflict of interest.}
The authors declare no conflict of interest.

\bigskip\bigskip\bigskip
%\newpage
\bibliographystyle{mybibstyle}
\bibliography{bibliografia}

\end{document}

%% file: Structure.tex
\usepackage[utf8]{inputenc}

\usepackage[top=3.5cm,bottom=3.5cm, left=4cm,right=4cm,   heightrounded,marginparwidth=1.5cm, marginparsep=1cm]{geometry} %for margins

\usepackage[shortlabels]{enumitem} % for lists

\usepackage[square,numbers,merge,comma,sort&compress]{natbib} % bibliography
\makeatletter
\def\NAT@spacechar{\,}  % define space inside [1,\,2]
\makeatother

\usepackage{amsmath,amssymb,amsfonts,amsthm,amsbsy}
\usepackage{mathtools} % for {rcases}

\usepackage{fnpct}% for multiple footnote at same point
\setfnpct{dont-mess-around} % no other kerning and punctuation switching

\usepackage{slashed,cancel,latexsym}
\usepackage{old-arrows,comment,relsize,setspace,moresize}
\usepackage{epsfig}
\usepackage{changepage,adjustbox}
\usepackage{mathrsfs,calligra,aurical,bold-extra} % fonts, \mathscr ...
\usepackage{calc,float,appendix}
\usepackage{bigints,xargs,extarrows}
\usepackage{empheq,soul} % for \hl{...}

\usepackage[svgnames]{xcolor}  % Load BEFORE tikz!
\definecolor{Blue}{rgb}{0.0, 0.0, 0.37}
\definecolor{Green}{rgb}{0.05, 0.45, 0.25}
\xdefinecolor{dogwoodrose}{rgb}{0.8, 0.1, 0.55}
\xdefinecolor{RRed}{rgb}{0.7, 0.1, 0.525}

\usepackage{tikz}
\usetikzlibrary{scopes,decorations.pathmorphing,patterns,calc,arrows,shapes.geometric,shapes.arrows,decorations.markings,plotmarks}
\usepackage{pgfplots}

\usepackage[blocks]{authblk}  % authors and affiliations

\usepackage{graphicx} % Required for including images
\graphicspath{{Figures/}} % Set the default folder for images

\usepackage[labelsep=colon]{caption}  % captions
\usepackage[labelsep=colon,aboveskip=10pt,belowskip=10pt]{subcaption}
\DeclareCaptionSubType * [roman]{table}
\usepackage{array,tabularx}

\usepackage{titlesec}
\titleformat{\section}{\normalfont\fontsize{12.5}{12}\bfseries}{\thesection}{0.5em}{}%{\phantomsection}
\titleformat{\subsection}{\normalfont\fontsize{10.5}{10}\bfseries}{\thesubsection}{0.5em}{}%{\phantomsection}
\titleformat{\subsubsection}{\normalfont\normalsize\bfseries}{\thesubsubsection}{0.5em}{}%{\phantomsection}

\titlespacing*{\section}{0pt}%
                {4ex plus 1ex minus .5ex}{1.75ex plus .25ex minus .25ex}
\titlespacing*{\subsection}{0pt}%
                {3.5ex plus 1ex minus .5ex}{1.25ex plus .2ex minus .2ex}
\titlespacing*{\subsubsection}{0pt}%
                {2.5ex plus 0.75ex minus .2ex}{0.75ex plus .15ex minus .15ex}
\titlespacing*{\paragraph}{0pt}%
                {1.85ex plus 0.5ex minus .15ex}{1em}

\usepackage{titletoc}
\addtocontents{toc}{\addvspace{-0.75em}}  % reduce vert space before TOC
\usepackage{bm}  % \mathbbm only has digits "1" and "2"
\usepackage{dsfont}  % double math characters (but only digit "1")

\DeclareMathAlphabet{\mathpzc}{OT1}{pzc}{m}{it}
\DeclareMathAlphabet{\mathcal}{OMS}{cmsy}{m}{n}
\DeclareSymbolFontAlphabet{\Scr}{rsfs}
\DeclareMathAlphabet{\mathbold}{U}{BOONDOX-ds}{m}{n}
\SetMathAlphabet{\mathbold}{bold}{U}{BOONDOX-ds}{b}{n}
\DeclareMathAlphabet{\mathcalboondox}{U}{BOONDOX-calo}{m}{n}
\SetMathAlphabet{\mathcalboondox}{bold}{U}{BOONDOX-calo}{b}{n}
\DeclareMathAlphabet{\mathbcalboondox}{U}{BOONDOX-calo}{b}{n}

\DeclareFontFamily{U}{matha}{\hyphenchar\font45}
\DeclareFontShape{U}{matha}{m}{n}{ <5> <6> <7> <8> <9> <10> gen * matha
                    <10.95> matha10 <12> <14.4> <17.28> <20.74> <24.88> matha12}{}
\DeclareSymbolFont{matha}{U}{matha}{m}{n}
\DeclareMathSymbol{\varleftarrow}{3}{matha}{"D0}
\DeclareMathSymbol{\varrightarrow}{3}{matha}{"D1}
\DeclareMathSymbol{\simeq}{3}{matha}{"14}
\DeclareMathSymbol{\sim}{3}{matha}{"12}
\DeclareMathSymbol{\ll}{3}{matha}{"21}
\DeclareMathSymbol{\gtrsim}{3}{matha}{"C1}
\DeclareMathSymbol{\lesssim}{3}{matha}{"C0}

\newcommand\linkcol{RRed}

\usepackage[breaklinks=true,backref=page]{hyperref}
\hypersetup{
    bookmarks=true,         % show bookmarks bar?
    bookmarksnumbered=true,
    pdftoolbar=true,        % show Acrobat’s toolbar?
    pdfmenubar=true,        % show Acrobat’s menu?
    pdffitwindow=false,     % window fit to page when opened
    pdfauthor={},     % author
    pdfsubject={},   % subject of the document
    pdfcreator={},   % creator of the document
    pdfproducer={},  % producer of the document
    pdfpagemode={UseNone},
    pdfstartview={FitH},
    colorlinks=true,
    plainpages,
    linktoc=page,
    citecolor=blue,
    filecolor=black,
    linkcolor=\linkcol,
    urlcolor=Grey,
}
\renewcommand*{\backref}[1]{}
\renewcommand*{\backrefalt}[4]{%
\ifcase #1 %
\relax
\or
~{\small [\textsc{p.~\fns{\!#2}}]}
\else
~{\small [\textsc{p.~\fns{\!#2}}]}%
\fi}

\usepackage{footnotebackref}
\usepackage{hypernat} % Load it HERE, after NATBIB and HYPERREF
\usepackage{cleveref} % Load it HERE, after HYPERREF

\Crefname{figure}{Fig.}{Figs.}

\makeatletter
\normalsize
\makeatother

\makeatletter
\g@addto@macro\bfseries{\boldmath}
\makeatother

\newcommand\fns{\footnotesize}

\newcommandx\Hodge[1][1=4,usedefault]{{}^{\star_{#1}}}

\newcommand\vF{v_{\textsc{f}}}
\newcommand\I{\mathcal{I}}

\newcommand\PP{\mathcal{P}}

\newcommand\LDOS{\rho_{{}_\textsc{s}}}%{\rho_{{}_\textsc{dos}}}
\newcommand\EF{E_{\textsc{f}}}
\newcommand\zetaN{\zeta_{{}_{N}}}

\providecommand{\abs}[1]{\left\lvert#1\right\rvert}

\newcommandx{\overbar}[1]{\mkern
1.5mu\overline{\mkern-2.0mu#1\mkern-2.0mu}\mkern 1.5mu}
\newcommandx{\overbarcal}[1]{\mkern                   6.0mu\overline{\mkern-5.5mu#1\mkern-1.0mu}\mkern 1.5mu}